\documentclass[twocolumn]{aastex6}

\begin{document}
\title{Experimental determination of whistler wave dispersion relation in the solar wind}
\author{D. Stansby, T. S. Horbury,  C. H. K. Chen, and L. Matteini}
\affil{Department of Physics, Imperial College London, London, SW7 2AZ, United Kingdom}
\email{david.stansby14@imperial.ac.uk}
\shortauthors{Stansby et. al.}
\shorttitle{Experimental whistler wave dispersion}

\begin{abstract}
The origins and properties of large amplitude whistler wave packets in the solar wind are still unclear. In this Letter we utilise single spacecraft electric and magnetic field waveform measurements \added{from the ARTEMIS mission} to calculate the plasma frame frequency and wavevector of individual wave packets over multiple intervals. This allows direct comparison of experimental measurements with theoretical dispersion relations to identify the observed waves as whistler waves. The whistlers are right-hand circularly polarised, travel anti-sunward and are aligned with the background magnetic field. Their dispersion is strongly affected by the local electron parallel beta in agreement with linear theory. The properties measured are consistent with the electron heat flux instability acting in the solar wind to generate these waves.
\end{abstract}

\section{Introduction}
Whistler waves are right-hand polarised plasma waves with frequencies between the ion and electron gyrofrequencies. They have been observed in the heliosphere using spectral \citep{Neubauer1977, Kennel1980, Coroniti1982, Lengyel-Frey1994, Lengyel-Frey1996, Lacombe2014}, magnetic field waveform \citep{Moullard2001, Wilson2009, RamirezVelez2012, WilsonIII2013}, and electric field waveform \citep{Breneman2010} measurements, and were identified based on spacecraft frame observations of their frequency and polarisation. With single spacecraft measurements of either the electric or magnetic field this is the only accessible information about the waves, and a dispersion relation must be assumed to calculate plasma frame frequencies and wavevectors. However, with simultaneous electric and magnetic field observations, frequencies and wavevectors can be measured independently, transformed into any inertial frame of reference, and an experimental dispersion relation can be determined. Measuring these properties for whistler waves is important to help determine how they are generated, and once generated how they interact with other waves and particles in the solar wind. 

There are several instabilities that can create plasma waves in the whistler wave frequency range: the electron firehose instability, whistler anisotropy instability, and whistler heat flux instability \citep{Gary2005}. The free energy to drive these instabilities comes from non-Maxwellian electron distribution functions, which in the solar wind consist of a dense core, a suprathermal halo, and a magnetic field aligned anti-sunward travelling strahl \citep{Pilipp1987, Stverak2009}. Because each instability is activated by characteristic distribution functions, observations of electron distributions can help identify active instabilities. For example, \cite{Lacombe2014} showed that observed electron distribution functions in the solar wind were sometimes unstable to the electron heat flux instability (caused by the electron strahl) when whistlers were observed. In addition, \cite{Moullard2001} showed examples of enhanced strahl number densities when whistlers were observed. Each instability also generates waves at characteristic wavevectors, frequencies, and polarisations, which can be used to determine the source instability. \cite{Zhang1998} showed that whistler waves in the solar wind travel predominantly anti-sunward, which is expected if the anti-sunward travelling electron strahl causes a heat flux instability. Further characterisation of the observed whistler waves is possible, which can provide more evidence for their origin and the active instabilities in the solar wind.

Regardless of their source, whistler waves will undergo wave-particle interactions. They are present at least 10\% of the time in the solar wind \citep{Lacombe2014}, so could play an important role in the global transfer of energy from fields to particles. For example, whistler wave interactions have a central role in theories seeking to explain the observed scattering of electrons from the strahl to the halo \cite[e.g.,][]{Saito2007, Vocks2011, Seough2015}.  Predicting which part of the distribution function whistler waves will interact with is important to constrain theories concerning wave-particle interactions, and requires knowledge of both wavevector and frequency.

In this Letter we calculate plasma frame properties of individual wavepackets detected in the whistler wave frequency range across multiple intervals using single spacecraft electric and magnetic field observations (Section \ref{sec:processing}). This allows us to present an experimental dispersion relation for these waves in the solar wind for the first time, and confirm their identification as whistler waves (Section \ref{sec:results}). We also discuss the implications for their generation and subsequent wave-particle interactions (Section \ref{sec:discussion}).

\section{Data set}
Data from the ARTEMIS mission \citep{Angelopoulos2011} is used in this study. The FGM instrument \citep{Auster2008} measures the 3D magnetic field and is used to determine the background magnetic field, $\mathbf{B}_{0}$. The SCM instrument \citep{Roux2008} also measures the 3D magnetic field and provides a reliable AC measurement above $\sim$4 Hz. This is used to determine the fluctuating magnetic field, $\delta \mathbf{B}$.  The EFI instrument \citep{Bonnell2008} measures the 3D electric field. The spin axis measurement is less accurate than the spin plane measurements, so has not been used here. The spin plane components have been used to determine the 2D fluctuating electric field, $\delta \mathbf{E}$, using the method described in Section \ref{sec:E field processing}. The ESA instrument \citep{McFadden2008a} measures both ion and electron distribution functions. From these distributions ground calculated moments were used for the solar wind bulk velocity ($\mathbf{v}_{sw}$), ion number density ($n_{i}$), and the electron temperatures perpendicular ($T_{e\perp}$) and parallel ($T_{e\parallel}$) to $\mathbf{B}_{0}$. The bulk velocity and ion number density were corrected assuming an alpha to proton number density ratio of $n_{\alpha} / n_{p} = 0.04$ as detailed in \cite{McFadden2008}. To avoid problems with spacecraft potential effects $n_{e}=n_{p}+2n_{\alpha}$ was used as a best estimate of the electron number density.

\subsection{Wavepacket selection}
\label{sec:wavepacket selection}
We have identified 7 intervals (listed in Table \ref{tab:intervals}) during which ARTEMIS probes P1 or P2 were in the solar wind, showed no evidence of magnetic connection to either the Earth's bow shock or the Moon, were in particle burst mode, and showed evidence of large amplitude magnetic field fluctuations above the background turbulence level. In each interval the electric and magnetic field were measured at 128 samples/second and full particle distributions and their associated moments measured every 3 seconds.

\begin{table}
	\centering
	\begin{tabular}{| c | c | c | c |}
		\tableline
		Probe	& Date		& Start Time (UT)	& End Time (UT)	\\ \tableline \tableline
		P1 		& 2010 Oct 8	& 00:11:18		& 00:21:15		\\ \tableline
		P1 		& 2010 Oct 8	& 00:22:58 		& 00:32:55		\\ \tableline
		P1 		& 2010 Oct 8	& 00:55:13		& 01:05:03		\\ \tableline
		P1		& 2010 Oct 8	& 04:56:10		& 05:06:47		\\ \tableline \tableline
		P2	 	& 2010 Nov 9	& 10:11:34		& 10:21:51		\\ \tableline
		P2		& 2010 Nov 9	& 10:47:38		& 10:56:27		\\ \tableline \tableline
		P2	 	& 2011 May 9	&16:32:19			& 16:43:00		\\ \tableline
	\end{tabular}
	\caption{Selected intervals used in this letter.\label{tab:intervals}}
\end{table}

To automatically detect individual wavepackets in the SCM magnetic field data, we used a similar method to \cite{Boardsen2015}. A Morlet trace power spectrogram \citep{Torrence1998} was calculated in the frequency range 4 Hz - 64 Hz, and the average power over the whole interval taken. All data points over 4 times the average power were marked, and connected component labelling used to select connected islands in the spectrogram containing more than 512 points. The earliest and latest time in each island determined the start and end of each wavepacket. The lowest and highest frequencies in each island determined the lower and upper limits for band pass filtering electric and magnetic field data. Data within each wavepacket was then processed as described in the following section.

\section{Data processing}
\label{sec:processing}
\subsection{Electric field}
\label{sec:E field processing}
In the solar wind the body of the spacecraft provides a barrier to the bulk flow. Directly downstream of the spacecraft a wake is formed that contains large electric fields, which dominate the signal and are measured each time one of the EFI probes enters the wake. The wake shows up as large discontinuous jumps in the time series making it possible to automatically detect and remove these periods. Approximately 30\% of the data points are removed by this process.

The other interference comes from the spin of the spacecraft, which introduces a complex large amplitude signal that repeats itself every spin period. To remove this the time series around each wavepacket was divided into segments, each a spin period long, and the average segment shape calculated over 12 spin periods ($\sim$36 seconds). This spin period average was then subtracted from each segment individually. This removed the low frequency-high amplitude spin tone whilst preserving the high frequency-low amplitude wave signal within the wavepacket.

The electric field was then Lorentz transformed into the solar wind bulk velocity frame. Because the observed waves have large phase speeds compared to $\left |\mathbf{v}_{sw}\right |$ this was a small correction. Finally each segment was individually band pass filtered with a 1st order Butterworth filter, using the frequencies found when selecting each wavepacket in Section  \ref{sec:wavepacket selection}. The top panel of Figure \ref{fig:wavepacket} shows an example electric field signal after processing, with gaps each time a wake spike has been removed.

\subsection{Wavepackets}
For each wavepacket the plasma frame frequency ($\omega$), wavevector ($\mathbf{k}$), and polarisation were calculated as follows. The spacecraft frame frequency ($\omega_{0}$) was taken as the frequency within the wavepacket at which the trace power spectrogram was a maximum. It is related to the plasma frame frequency, wavevector, and solar wind bulk velocity via.
\begin{equation}
	\omega_{0}=\omega+\mathbf{k}\cdot\mathbf{v}_{sw}
	\label{eq:Taylor}
\end{equation}
To determine $\hat{\mathbf{k}}$, minimum variance analysis was used on $\delta\mathbf{B}$ to determine the normal vector to the plane in which the fluctuations lay \citep{Sonnerup1967}. Using this method results in a $180^{\circ}$ ambiguity in $\hat{\mathbf{k}}$. This was resolved by calculating the Poynting vector, which is parallel to the wavevector. With only two components of the electric field only one component of the Poynting vector could be calculated, which was enough to determine the hemisphere in which $\hat{\mathbf{k}}$ lay and resolve the $180^{\circ}$ ambiguity.

This leaves two unknowns in Equation \ref{eq:Taylor}, $\omega$ and $\left| \mathbf{k}\right|$. The ratio of these two quantities is the phase speed of the wave. \replaced{For purely electromagnetic waves (such as whistler waves) with both $\delta \mathbf{B}$ and $\delta\mathbf{E}$ perpendicular to $\mathbf{k}$, the phase speed is related to the electric and magnetic field magnitudes by}{For whistler waves propagating parallel to $\mathbf{B}_{0}$, $\delta\mathbf{E}$ is perpendicular to $\mathbf{k}$ \citep{Tokar1985}, and the phase speed is related to the electric and magnetic field magnitudes by}
\begin{equation}
	v_{p}\equiv\frac{\omega}{\left|\mathbf{k}\right|}=\frac{\left|\mathbf{\delta E}\right|}{\left|\mathbf{\delta B}\right|} 
	\label{eq:Phase speed}
\end{equation}
With only two components of the electric field, $\left|\delta\mathbf{E}\right|$ could not be fully evaluated. As long as a wave is elliptically polarised at most one field component is always zero, and the phase speed can be calculated using
\begin{equation}
	v_{p}=\left\langle\frac{\sqrt{\delta E_{x}^{2}+\delta E_{y}^{2}}}{\sqrt{\delta B_{x}^{2}+\delta B_{y}^{2}}}\right\rangle
	\label{eq:2dphasespeed}
\end{equation}
with the two components measured in the spacecraft spin plane. The average was taken over multiple wave periods, ignoring the highest 10\% and lowest 10\% of single point measurements to remove anomalously large values due to simultaneously low $\delta B_{x}$ and $\delta B_{y}$ measurements. Equations \ref{eq:Taylor} and \ref{eq:2dphasespeed} along with minimum variance analysis allowed $\omega$ and $\mathbf{k}$ to be uniquely determined.

The ellipticity of the wave was calculated from the minimum variance eigenvalues \citep[see][for details]{Born1999}. The sign of the ellipticity gives the spacecraft frame polarisation which was converted to a plasma frame polarisation using the solar wind bulk velocity, plasma frame frequency, and wavevector.

At this point two quality checks were imposed on each wave packet:
\begin{itemize}
	\item  Only wavepackets whose maximum and minimum variance eigenvalues satisfied $a_{max} / a_{min} > 10$ were kept. This selected for plane polarised waves, but did not select between linear or circular polarisation. (108 wavepackets failed this test)
	\item Only wavepackets where over 60\% of the Poynting flux z component measurements had the same sign were kept.  This ensured a reliable determination of $\hat{\mathbf{k}}$. (141 wavepackets failed this test)
\end{itemize}
This left 289 individual wavepackets, each with calculated plasma frame properties.

\begin{figure}
	\epsscale{1.2}
	\plotone{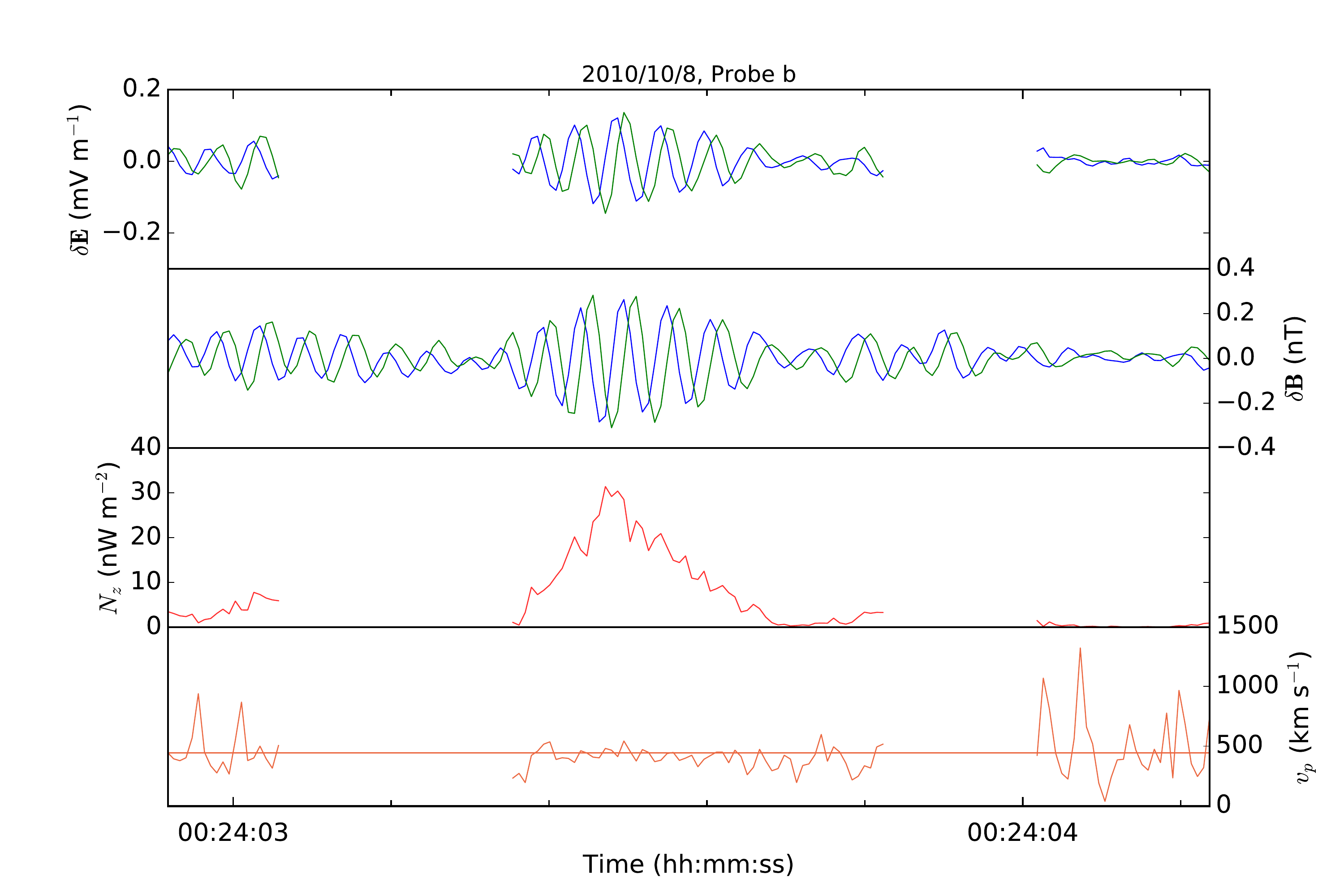}
	\caption{Time series of $\delta\mathbf{E}$ spin plane components (top panel), corresponding components of $\delta\mathbf{B}$ (second panel), component of Poynting flux perpendicular to measured components of $\delta\mathbf{E}$ (third panel), and phase speed (bottom panel). The horizontal line shows the average phase speed ignoring the lowest and highest 10\% of single point measurements.}
	\label{fig:wavepacket}
\end{figure}

Figure \ref{fig:wavepacket} shows an example of filtered data for a single detected wavepacket. The spin plane electric and magnetic fields show a similar form as expected. The component of the Poynting flux perpendicular to the spin plane is strongly enhanced at times where there is a visible wave packet. Although the phase speed is sensitive to small variations in $\delta\mathbf{E}$ and $\delta\mathbf{B}$ it maintains a steady mean value of $\sim$500 km s$^{-1}$ during the wavepacket.

\section{Results}
\label{sec:results}
\begin{figure}
	\plotone{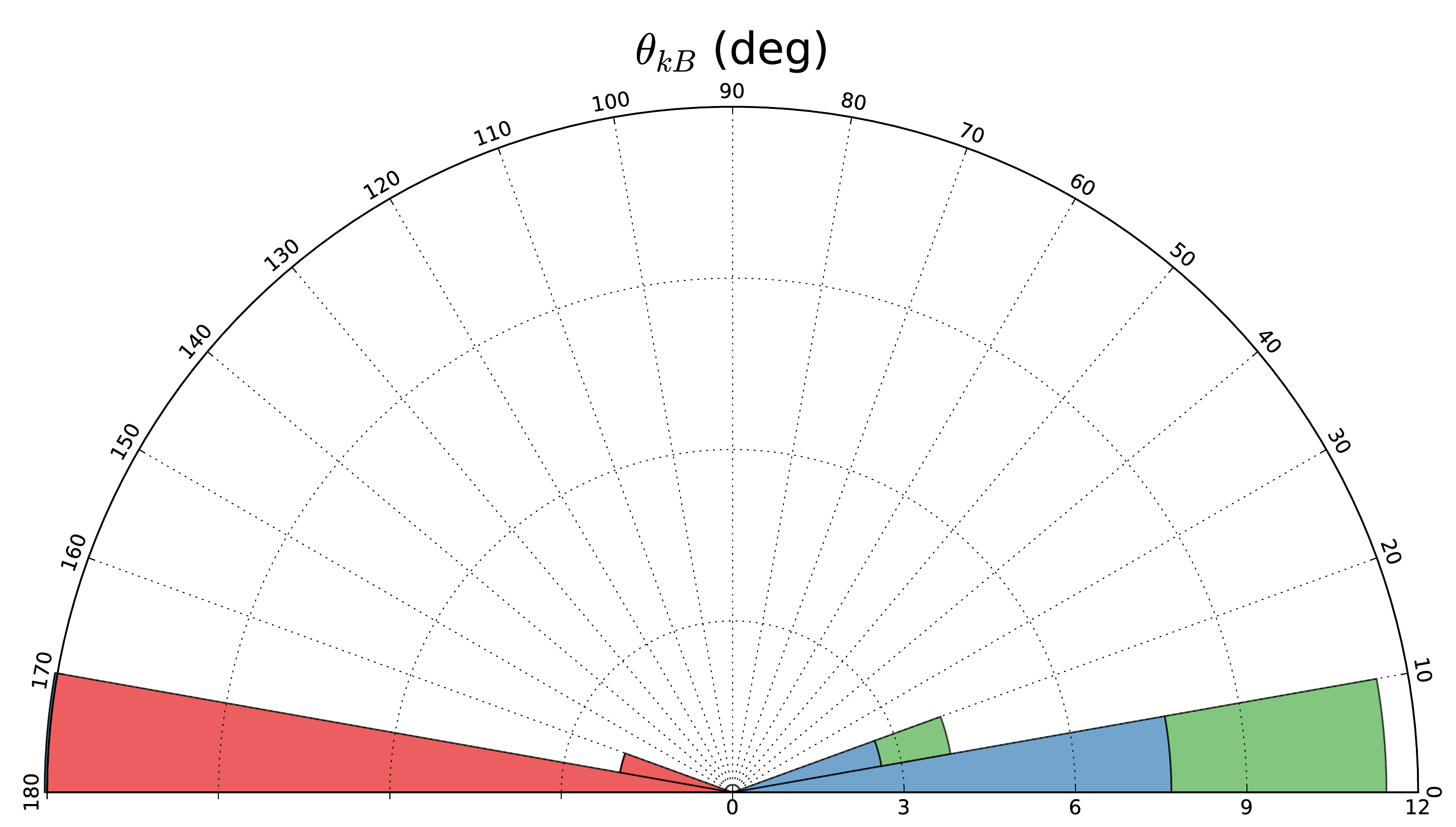}
	\plotone{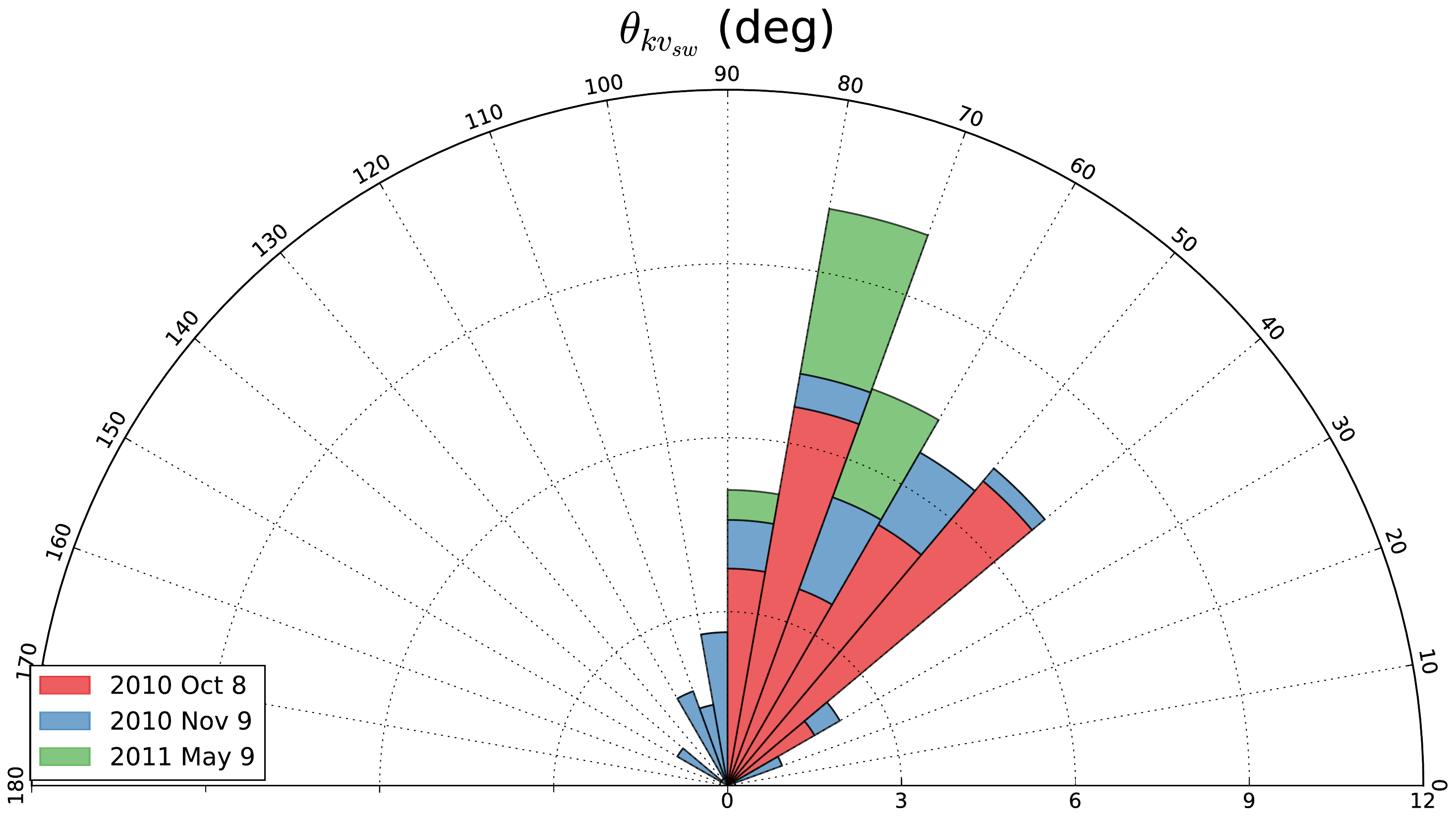}
	\caption{Polar histograms of angles between $\mathbf{k}$ and $\mathbf{B}_{0}$ (top panel) and $\mathbf{k}$ and $\mathbf{v}_{sw}$ (bottom panel). Bins are \added{stacked and} coloured by the day on which they were observed. The number of wavepackets in each bin is proportional to the bin area.}
	\label{fig:histograms}
\end{figure}

Figure \ref{fig:histograms} shows polar histograms of the angles between $\mathbf{k}$ and $\mathbf{B}_{0}$ (top panel), and $\mathbf{k}$ and $\mathbf{v}_{sw}$ (bottom panel). 98\% of the waves travelled anti-sunward and all travelled within 20$^{\circ}$ of $\mathbf{B}_{0}$ or $-\mathbf{B}_{0}$, consistent with \cite{Zhang1998}.

In order to identify the wave mode of these fluctuations, experimental data were compared to theoretical dispersion relations computed using the Waves in Homogeneous, Anisotropic Multicomponent Plasmas (WHAMP) linear dispersion solver  \citep{Roennmark1982}. A two component proton-electron plasma was used with each species having a non-drifting bi-Maxwellian distribution function. To match typical solar wind conditions we set $n_{p}=n_{e}=5$ cm$^{-3}$,  $B_{0}$ = 5 nT, and parallel and perpendicular temperatures were set by specifying the parallel beta, $\beta_{\parallel}=2\mu_{0}nk_{B}T_{\parallel}/B_{0}^{2}$, and temperature anisotropy, $\mathcal{A}=T_{\perp}/T_{\parallel}$, for each component. Frequencies and wavenumbers were normalised to local plasma scales: the electron gyro frequency $\Omega_{ce}=q_{e} B_{0}/m_{e}$ and the electron gyro radius $\rho_{e}=v_{th,e}/\Omega_{ce}$, where the electron thermal speed is $v_{th,e}=\sqrt{2k_{B}T_{\perp e}/m_{e}}$. Normalising to electron scales ensured that neither the proton beta nor proton temperature anisotropy affected the normalised frequency or wavevector. Within this model the only wave mode predicted to propagate at the observed frequencies is the whistler wave. Variations in proton parameters do not significantly alter whistler wave dispersion curves, so we set $\beta_{p\parallel}=1$ and $T_{p\perp}/T_{p\parallel}=1$ for all calculations.  Because observationally all waves travel along the background magnetic field, only the wavenumber ($k\equiv\left|\mathbf{k}\right|$) is plotted and all dispersion curves are for propagation parallel to $\mathbf{B}_{0}$.
 
\begin{figure}
	\plotone{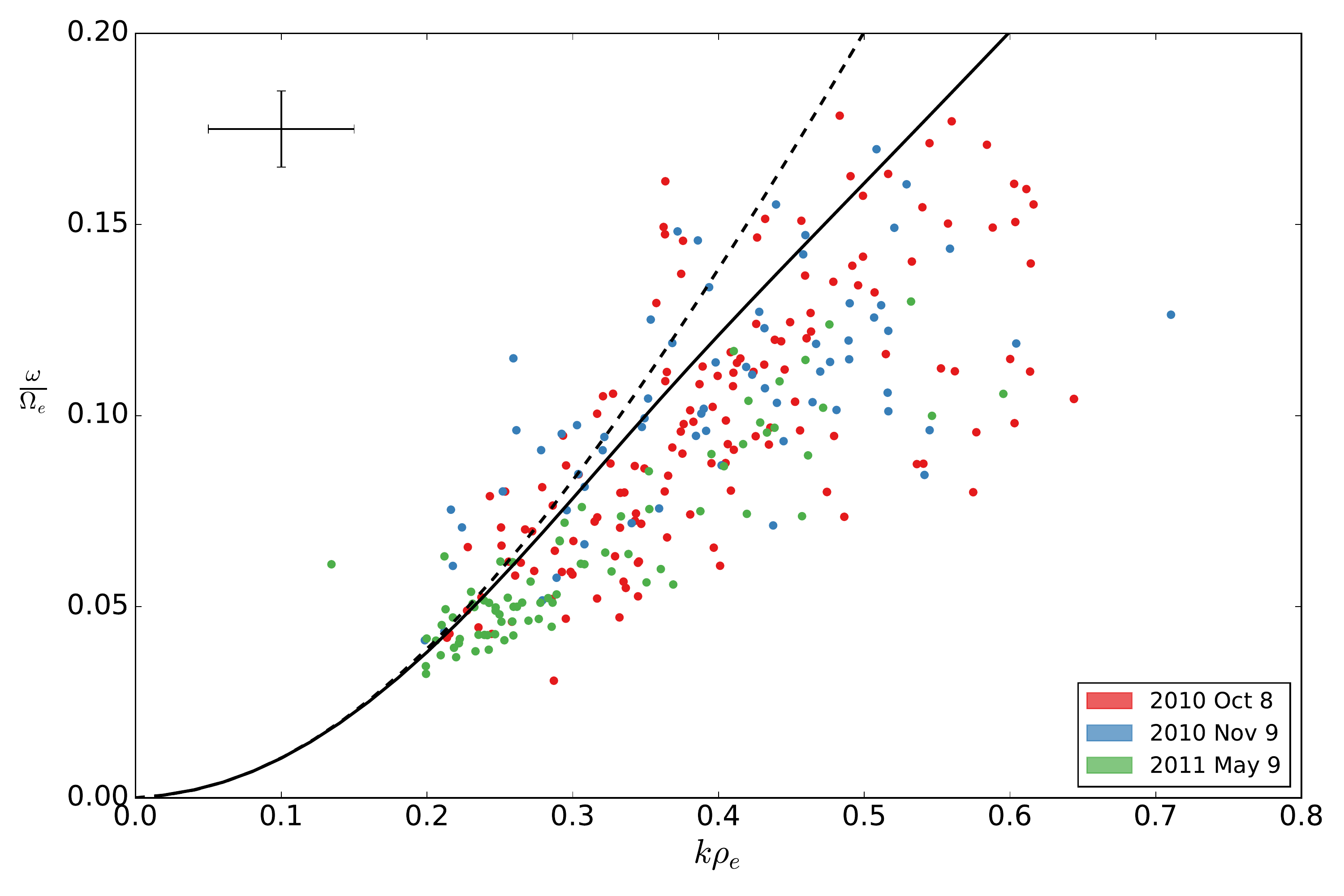}
	\caption{Experimental dispersion relation. The black dashed line shows the cold whistler dispersion relation and the black solid line shows the $\beta_{e}=1$, $T_{e\perp}/T_{e\parallel}=1$ whistler dispersion relation. Points are coloured by the day on which they were observed. A typical error bar is shown in the top left.}
	\label{fig:dispersion}
\end{figure}

Figure \ref{fig:dispersion} displays a scatter plot of measured frequencies and wavenumbers, coloured by the day on which they were observed to compare different solar wind conditions. A typical error bar is shown in the top left, calculated from uncertainties in measuring the wave frequency and phase speed; this shows the spread of the data cannot be solely attributed to experimental error. The black dashed line shows the cold whistler dispersion relation \citep{Stix1992} and the black solid line the warm whistler dispersion for $\beta_{e\parallel}=1$ and $T_{e\perp}/ T_{e\parallel}=1$. Points measured on different days by different probes follow the same trend, clustered around the whistler wave dispersion relations. At higher wavenumbers the dispersion curves diverge, with the warm dispersion relation staying closest to the centre of spread. Adding a strahl-like electron beam to the Maxwellian core would lower the warm dispersion curve slightly \citep{Gary2005}, providing a closer fit to the centre of spread. The distribution of wave ellipticity in the plasma frame (not shown here) is strongly peaked at +1, meaning the observed waves are RH circularly polarised. Both this and agreement with theoretical dispersion relations confirms the identification of these wavepackets as whistler waves.

\begin{figure}
	\plotone{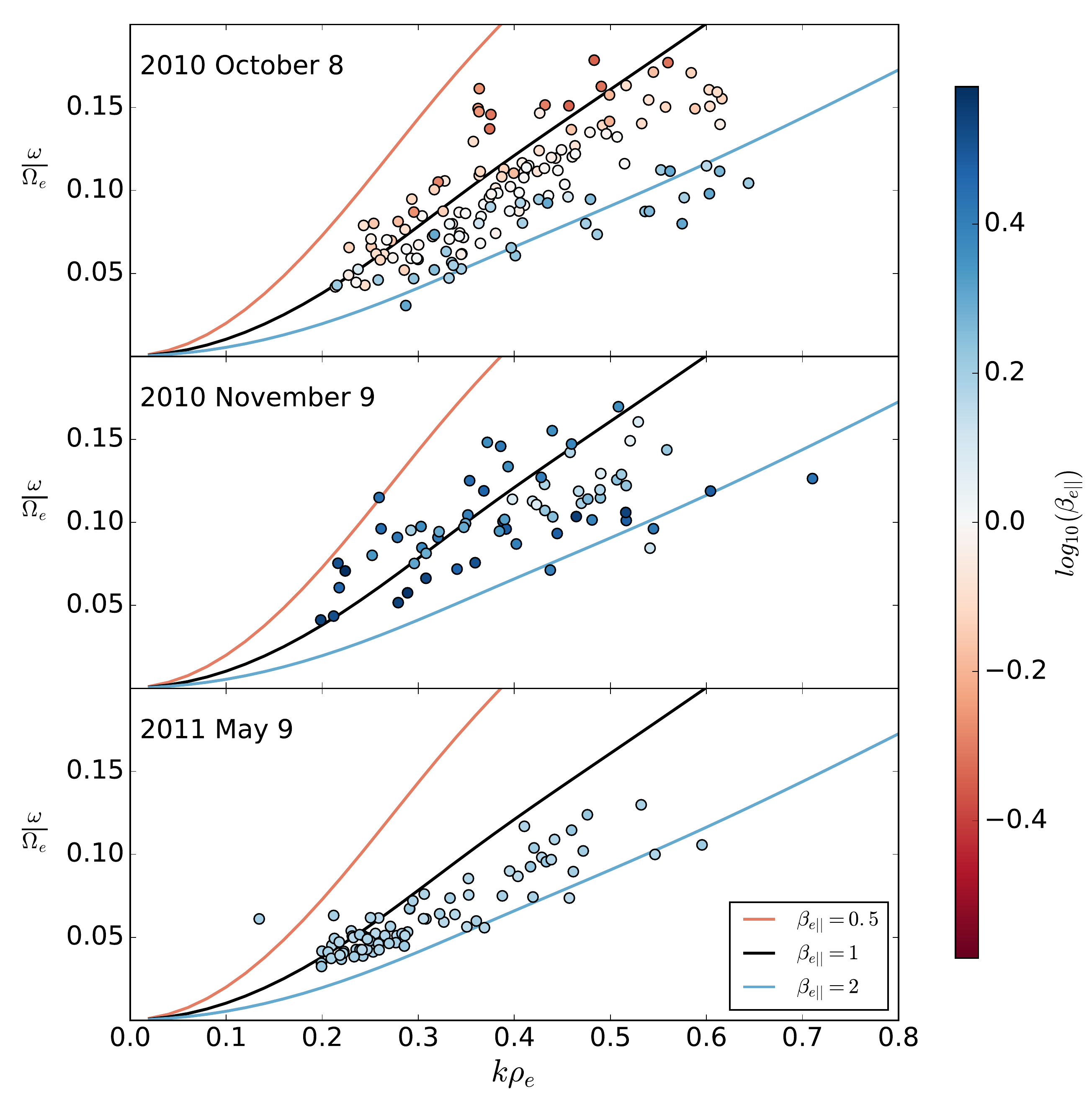}
	\caption{Experimental dispersion relation with points coloured by $\beta_{e}$. Different panels correspond to different dates of observation. Overplotted lines show the dispersion relation for $T_{e\perp}/T_{e\parallel} = 1$ and varying electron beta.}
	\label{fig:anibeta}
\end{figure}

To investigate the cause of the scatter in Figure \ref{fig:dispersion}, we looked at how the dispersion depends on the electron beta and temperature anisotropy. The range of $T_{e\perp}/T_{e\parallel}$ observed in our data is 0.83 - 1.03, which is typical for the solar wind \citep{??Tver??K2008}. These variations are not large enough to significantly alter the dispersion of whistler waves. In contrast the range of $\beta_{e\parallel}$ observed can significantly alter the whistler wave dispersion. In Figure \ref{fig:anibeta}  the data are split into different observation days and coloured by the local parallel electron beta. Overplotted are warm whistler dispersion curves for $T_{e\perp}/T_{e\parallel}=1$ and different $\beta_{e\parallel}$ values.  In the first and third panels of Figure \ref{fig:anibeta} points with higher $\beta_{e\parallel}$ have a higher wavenumber at a fixed frequency, agreeing well with linear theory. For a given frequency the wavenumber of a wave may vary by as much as a factor of 2 for the range of $\beta_{e\parallel}$ observed. In contrast to the first and third panels, data in the second panel do not appear to agree with linear theory. All points here have large $\beta_{e\parallel}$ values, but lie on both sides of the $\beta_{e\parallel}=1$ dispersion relation as opposed to only below it. The large $\beta_{e\parallel}$ values could be caused additional non-Maxwellian features in the solar wind on this day. These features cannot be captured by the simple two component bi-Maxwellian model without drifts used here, which could explain the difference between the data and example dispersion relations.

\section{Discussion}
\label{sec:discussion}
Through the use of \added{single spacecraft} simultaneous electric and magnetic field measurements, we have constructed an experimental dispersion relation (Figure \ref{fig:dispersion}) to identify multiple large amplitude wavepackets in the solar wind as whistler waves, and considered the effect of the local plasma properties on their dispersion (Figure \ref{fig:anibeta}). For the range of plasma parameters observed in the solar wind the electron beta plays the largest role in determining the wavenumber of a whistler wave at a given frequency. Linear theory qualitatively agrees with our data when $0.5\lesssim\beta_{e\parallel}\lesssim 2$.

There are three instabilities which could produce waves at the range of wavenumbers observed: the electron firehose instability,  the whistler anisotropy instability, and the whistler heat flux instability \citep{Gary2005}. The electron firehose produces either non-propagating structures with $\omega=0$ or LH polarised waves \citep{Li2000b, Camporeale2008a}. This instability is ruled out as we have measured neither of these properties. The whistler anisotropy instability requires large temperature anisotropies. The largest anisotropy recorded in our dataset is $T_{e\perp}/T_{e\parallel} = 1.03$ which is not large enough to provide significant growth rates over the range of wavenumbers observed, so this instability is also ruled out. Additionally, the whistlers travelled preferentially at small angles to $\mathbf{B}_{0}$ and anti-sunward, the same direction as the electron strahl. These lines of evidence favour the hypothesis that they were generated by the heat flux instability, which has the highest growth rate at $\theta_{kB}=0$ and in the same direction as the electron heat flux \citep{Gary1975}. This result complements that of \cite{Lacombe2014}, who used particle data to show that in the presence of whistler waves the solar wind plasma was sometimes unstable to the heat flux instability, but not unstable to the electron firehose or whistler anisotropy instabilities.

Anti-sunward wave propagation has consequences for the allowable wave particle interactions. The resonance condition for whistler waves and electrons reads
\begin{equation}
	\omega-\Omega_{e}=k_{\parallel}v_{\parallel}
\end{equation}
where $v_{\parallel}$ is the velocity of the resonant particles parallel to $\mathbf{B}_{0}$ and $k_{\parallel}$ is the component of the wavevector along $\mathbf{B}_{0}$. Because $\omega<\Omega_{e}$ for all waves observed, $k_{\parallel}v_{\parallel}<0$ which means resonantly interacting waves and particles must be travelling in opposite directions. Once generated the observed waves cannot resonantly interact with the anti-sunward moving strahl.  The mean resonant velocity of our dataset is 2.7$v_{th,e}$ with an inter quartile range of 1.9$v_{th,e}$ - 3.3$v_{th,e}$, so these waves primarily interact with particles in the sunward halo, and could not perform the strahl scattering proposed in e.g. \cite{Vocks2005a} or \cite{Seough2015}. However, to observe whistler waves in this study their amplitude had to be significantly larger than that of the turbulent background, so we have not ruled out the presence of lower amplitude sunward travelling whistler waves.

\added{An experimental whistler dispersion relation in the solar wind has also recently been presented by \cite{Narita2016c}, who used used multi-spacecraft data from the MMS mission to measure the dispersion of broadband magnetic field turbulence. In contrast, here we have presented observations of an additional sporadic whistler population that exists on top of the background turbulence. The waves here propagate parallel to $\mathbf{B}_{0}$, whereas the waves presented by \cite{Narita2016c} propagate quasi-perpedicular to $\mathbf{B}_{0}$. Multi-spacecraft measurements with only magnetic field measurements can be used to measure the 4D wave power in a region of $(\omega, \mathbf{k})$ space determined by the spacecraft separation, whereas the method presented in this Letter is limited to measuring the dispersion of individual monochromatic waves. However, our method requires data from only a single spacecraft, which will be useful for the upcoming Solar Probe Plus and Solar Orbiter missions.}

Finally, we note that experimentally measured distribution functions can be used to predict the fastest growing wave mode and its properties \citep{Gary2015, Wicks2016, Jian2016}. Simultaneously observing the predicted waves and their properties using the method presented in this letter would provide strong evidence for in-situ plasma wave generation in the solar wind.

\acknowledgements
D. Stansby is supported by STFC studentship ST/N504336/1. T. S. Horbury and L. Matteini are supported by STFC grant  ST/N000692/1. C. H. K. Chen is supported by an Imperial College Junior Research Fellowship.

We acknowledge NASA contract NAS5-02099 and V. Angelopoulos for use of data from the THEMIS Mission. Specifically: C. W. Carlson and J. P. McFadden for use of ESA data, A. Roux and O. LeContel for use of SCM data, K. H. Glassmeier, U. Auster and W. Baumjohann for the use of FGM data, and J. W. Bonnell and F. S. Mozer for use of EFI data.
\software{WHAMP v.1.4.1 \citep{Roennmark1982}, Matplotlib v1.5.1 \citep{Droettboom2016}}


\listofchanges
\end{document}